Article

# COVID-19 Vaccine and Social Media in the U.S.: Exploring Emotions and Discussions on Twitter

Amir Karami [1,\*], Michael Zhu [2], Bailey Goldschmidt [3], Hannah R. Boyajieff [4] and Mahdi M. Najafabadi [5]

1. School of Information Science, University of South Carolina, Columbia, SC 29208, USA
2. Department of Psychology, University of South Carolina, Columbia, SC 29208, USA; mz3@email.sc.edu
3. College of Nursing, University of South Carolina, Columbia, SC 29208, USA; goldschb@email.sc.edu
4. Darla Moore School of Business, University of South Carolina, Columbia, SC 29208, USA; boyajieh@email.sc.edu
5. Graduate School of Public Health, City University of New York, New York, NY 10010, USA; Mahdi.Najafabadi@sph.cuny.edu
\* Correspondence: karami@sc.edu

**Abstract:** The understanding of the public response to COVID-19 vaccines is the key success factor to control the COVID-19 pandemic. To understand the public response, there is a need to explore public opinion. Traditional surveys are expensive and time-consuming, address limited health topics, and obtain small-scale data. Twitter can provide a great opportunity to understand public opinion regarding COVID-19 vaccines. The current study proposes an approach using computational and human coding methods to collect and analyze a large number of tweets to provide a wider perspective on the COVID-19 vaccine. This study identifies the sentiment of tweets using a machine learning rule-based approach, discovers major topics, explores temporal trend and compares topics of negative and non-negative tweets using statistical tests, and discloses top topics of tweets having negative and non-negative sentiment. Our findings show that the negative sentiment regarding the COVID-19 vaccine had a decreasing trend between November 2020 and February 2021. We found Twitter users have discussed a wide range of topics from vaccination sites to the 2020 U.S. election between November 2020 and February 2021. The findings show that there was a significant difference between tweets having negative and non-negative sentiment regarding the weight of most topics. Our results also indicate that the negative and non-negative tweets had different topic priorities and focuses. This research illustrates that Twitter data can be used to explore public opinion regarding the COVID-19 vaccine.

## 1. Introduction

By mid-May 2021, globally more than 158 million COVID-19 cases have been confirmed, and the death toll has reached more than 3.2 million [1]. In the U.S., more than 32 million cases and 580,000 deaths were reported [1]. During the COVID-19 pandemic, people were encouraged or enforced to stay at home, practice social distancing, and perform most of their work and life activities (e.g., shopping) remotely. This isolation led to an increase in using social media to receive news and express opinions [2]. Twitter users have posted more than 638 million COVID-19 related tweets between 1 January 2020, and 8 May 2020 [3]. Not surprisingly, #COVID19 and its variations were the top Twitter hashtags in 2020 and also in 2021 so far [4]. Not only citizens but also government officials have utilized Twitter to regularly share policies and news related to COVID-19 [5].

Several novel technologies have utilized computer power to study different aspects of the COVID-19 pandemic, its impacts, and how to respond to it efficiently and effectively [6,7]. A growing research trend argues that social media will play an important role in public health [8,9]. The first component of public health surveillance is monitoring, identifying, and evaluating health issues [10]. Social media offers a largely untapped opportunity for the first component. In public health surveillance, social media can help to provide communication in real-time and at relatively low cost [11], monitor public response to health issues [11], track disease outbreaks [12] and infectious disease [13], detect misinformation [14,15], identify target areas for intervention efforts [16], and disseminate pertinent health information to targeted communities [17].

A large number of studies have utilized Twitter data to understand public discussions around the COVID-19 pandemic. These studies have investigated a wide range of topics related to COVID-19, such as understanding public communication using qualitative content analysis [18], word frequency analysis and sentiment analysis [19], and topic modeling [20–22], examining misinformation [23], and measuring social distancing [24].

Low vaccine acceptance has a negative impact on herd immunity [25]. Therefore, public response to COVID-19 vaccines is a key success factor for developing nonpharmaceutical interventions and controlling the COVID-19 pandemic [26,27]. Traditional surveys are expensive and time-consuming, address limited health topics, and obtain small-scale data. To understand public opinion, we need a way to look beyond the individuals. Twitter can provide a great opportunity to understand public opinion regarding COVID-19 vaccines.

It is not the first time that people discussed vaccine issues on Twitter. Those vaccine discussions were used to identify hesitant communities regarding measles, mumps, and rubella combination (MMR), tetanus, diphtheria, pertussis (Tdap), and human papillomavirus (HPV) vaccines [28], analyze vaccine images [29], understanding the vaccine debate of Russian trolls [30], vaccine hesitancy [31], and sentiment analysis of HPV-related tweets [32]. Since 2020, some studies have utilized Twitter data to understand different issues related to the COVID-19 vaccine, such as exploring public opinion regarding vaccine hesitancy [33,34], vaccination in November 2020 [35], sentiment analysis of tweets posted between 1 March and 22 November 2020 [36], manual coding and content analysis of tweets posted between 1 and 22 November 2020 [37], misinformation [38,39], Spanish pro-vaccine campaign on Twitter between 14 and 28 December 2020 [40], anti-vaccination tweets posted between 1 January and 23 August, 2020 [41], and race-related discussions [42].

While the studies above offer valuable insights regarding COVID-19 vaccine issues, they have several limitations. First, the current research has not studied public opinion after public vaccination was kicked-off in November 2020. Second, there is no study on combining sentiment and semantic analysis to compare negative and non-negative tweets based on the weight of discussed topics.

The current study proposes an approach using computational and human coding methods to collect and analyze a large number of tweets to bridge the aforementioned gaps and provide a wider perspective on COVID-19 vaccine. This paper identifies the sentiment of tweets and their temporal trends, discovers major topics, and compares topics of negative and non-negative tweets. This paper addresses the following research questions:

1. How did the sentiment of tweets related to the COVID-19 vaccine change between November 2020 and February 2021?
2. What are the main topics in tweets related to the COVID-19 vaccine?
3. Is there a significant difference between topics in negative and non-negative tweets?
4. What are the top topics in negative and non-negative tweets?

This endeavor offers the following contributions. First, the proposed data analysis framework is a flexible approach that can be applied to other health issues. Secondly, three sentiment analyses are compared to identify the most efficient one. Third, this paper explores the sentiment of tweets during four months in 2020 and 2021. Fourth, we compare negative and non-negative tweets based on the weight of topics. Fifth, we identify top topics of negative and non-negative tweets.

## 2. Materials and Methods

This section provides details on the method used in this paper. We propose a methodology framework containing five components discussed below and depicted in Figure 1.

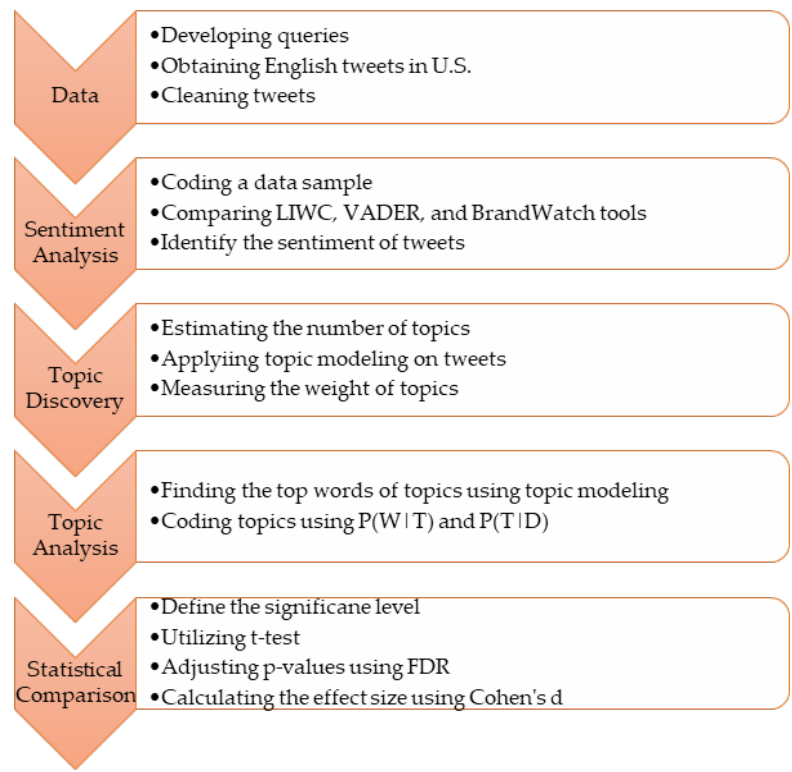

**Figure 1.** Research Framework.

*2.1. Data*

We collected data from a data service provider called BrandWatch. Due to the global COVID-19 pandemic, the dominant topic of social media was COVID-19. However, not all tweets were related to COVID-19. So, we limited our queries to Vaccine OR #Vaccine, OR Vaccination OR #Vaccination to collect 200,000 tweets (5000 tweets per every three days). These tweets were posted between 1 November 2020, and 28 February 2021. BrandWatch lets us collect English tweets posted in the United States by individual users (vs. other accounts such as organizations) and excludes retweets and comments posted by accounts related to spammers, businesses, marketing and advertising, pornographic, and automated accounts or bots. We have removed URLs, account names starting with @, and non-alphanumeric characters such as *. Then, we removed duplicate tweets from the same users and short tweets containing less than five terms. This process yielded 185,953 tweets.

*2.2. Sentiment Analysis*

This research focused on identifying negative and non-negative (positive and neutral) tweets. We utilized three methods offered by Linguistic Inquiry and Word Count (LIWC), Valence Aware Dictionary and sEntiment Reasoner (VADER), and Brandwatch. LIWC processes a document and counts the percentage of words that reflect positive and negative emotions using a group of built-in dictionaries [43]. For example, the word "cried" maps to negative emotion. We measured the difference between the positive emotion and negative emotion scores for each tweet. If the difference was less than zero, the sentiment of the tweet would be negative. Otherwise, the sentiment would be non-negative. VADER offers a rule-based sentiment analysis approach using machine learning [44]. VADER was developed in Python (https://github.com/cjhutto/vaderSentiment accessed on 20 May 2021) using a compound score (CS). A tweet has a negative sentiment if CS ≤ −0.05. Otherwise, the tweet has a non-negative sentiment. Brandwatch also uses a rule-based approach for sentiment analysis. This tool takes "all the words and phrases implying positive or negative sentiment and applies rules that consider how context might affect the tone of the content" [45]. Brandwatch categorizes tweets in to positive, negative, or neutral. We combined positive and neutral tweets as non-negative tweets.

To compare these three approaches, we randomly selected a sample of 1000 tweets. Each tweet was coded manually by two coders to show whether a tweet has a negative sentiment. Out of the coded data, we used the 719 tweets that the two coders had a full agreement on their sentiment. Then, we compared the 719 tweets coded by humans with the results of the three methods. The comparison showed that Brandwatch had the highest agreement (75.52%), with the human coding followed by VADER (66.89%) and LIWC (63.28%). Therefore, we utilized Brandwatch for our sentiment analysis in this study.

*2.3. Topic Discovery*

To identify topics discussed in the tweets, we utilized topic modeling. Latent Dirichlet Allocation (LDA) is an unsupervised text mining approach that identifies themes in a corpus [46,47]. This method has been applied to different corpora in different domains, including health and politics [10,48–50]. The outputs of LDA for *n* documents (tweets), m words, and t topics provided two matrices: the probability of each of the words per topic or $P(W_i|T_k)$ and the probability of each of the topics per document or $P(T_k|D_j)$ [51]. $P(W_i|T_k)$ recognizes related words representing a theme semantically, and $P(T_k|D_j)$ shows the weight of each topic per document.

$$\text{Words} \begin{bmatrix} P(W_1|T_1) & \cdots & P(W_1|T_t) \\ \vdots & \ddots & \vdots \\ P(W_m|T_1) & \cdots & P(W_m|T_t) \end{bmatrix} \overset{\text{Topics}}{} \quad \text{Words} \begin{bmatrix} P(W_1|T_1) & \cdots & P(W_1|T_t) \\ \vdots & \ddots & \vdots \\ P(W_m|T_1) & \cdots & P(W_m|T_t) \end{bmatrix} \overset{\text{Documents}}{}$$
$$P(W_i|T_k) \qquad\qquad\qquad P(T_k|D_j)$$

To estimate the number of topics, we used the C_V method [52] to measure the coherence for the number of topics from 2 to 100 topics. Developed in the Gensim Python package [52], the C_V method is highly correlated with human ratings [53]. We found the optimum number of topics at 26. Then, we applied the Mallet implementation of LDA [54] on our corpus. We set the Mallet at 26 topics and 4000 iterations. We also used the list of stop words in Mallet to remove the most common words such as "the," which do not provide semantic value for our analysis. We compare random five sets of 4000 iterations to validate the robustness of LDA. This experiment showed no significant difference between the mean and standard deviation of the log-likelihood of the five sets.

*2.4. Topic Analysis*

To understand the overall theme of topics, two coders qualitatively investigated each topic following two steps. First, the coders analyzed the top words of each topic using $P(W_i|T_k)$ and the top tweets of each topic using $P(T_k|D_j)$. The coders answered two questions: (Q1) "Does the topic have a meaningful theme?" (Q2) "Is the topic related to a vaccine?" The first step helped to remove the meaningless and irrelevant topics. In the second step, the coders used consensus coding [55] to identify a theme and create a label for each topic.

*2.5. Statistical Analysis*

We developed statistical tests to compare the average weight of topics in negative and non-negative tweets. First, we applied the two-sample t-test developed in the R 'mosaic' package [56]. Then, we defined the significance level based on the sample size [57] using $\frac{0.05}{\sqrt{\frac{N}{100}}}$ [58], where N is the number of tweets. As we had 185,953 tweets, the passing *p*-value was set at 0.001. Next, we adjusted *p*-values using the False Discovery Rate (FDR) method to minimize both false positives and false negatives [59].

Finally, to identify the magnitude of the difference between negative and non-negative tweets regarding the average weight of topics, we used the absolute effect size using Cohen's d calculated by dividing the mean difference by the pooled standard deviation [60]. We utilized the extended index classification, including very small (d = 0.01), small (d = 0.2), medium (d = 0.5), large (d = 0.8), very large (d = 1.2), and huge (d = 2.0) classifications [61]. This classification has two restrictions. First, it was developed based on small sample sizes $\leq 1000$ [62]. Second, Cohen's d is smaller in large samples [63]. To address these limitations, we obtained sample sizes used in developing the initial Cohen's d classification [62], including 8, 40, 60, 100, 200, 500, and 1000 random tweets. Then, we measured the mean of effect sizes of the sample sizes.

## 3. Results

We present our findings in three parts. The first part shows our sentiment analysis of tweets. After identifying the sentiment of tweets using the Brandwatch tool, we measured the frequency of negative and non-negative tweets. Figure 2 shows the rate of negative and non-negative tweets per month from November 2020 to February 2021. In total, 33.64% and 66.36% of tweets were negative and non-negative, respectively. The rate of negative tweets has a decreasing trend, and the rate of non-negative tweets has an increasing trend.

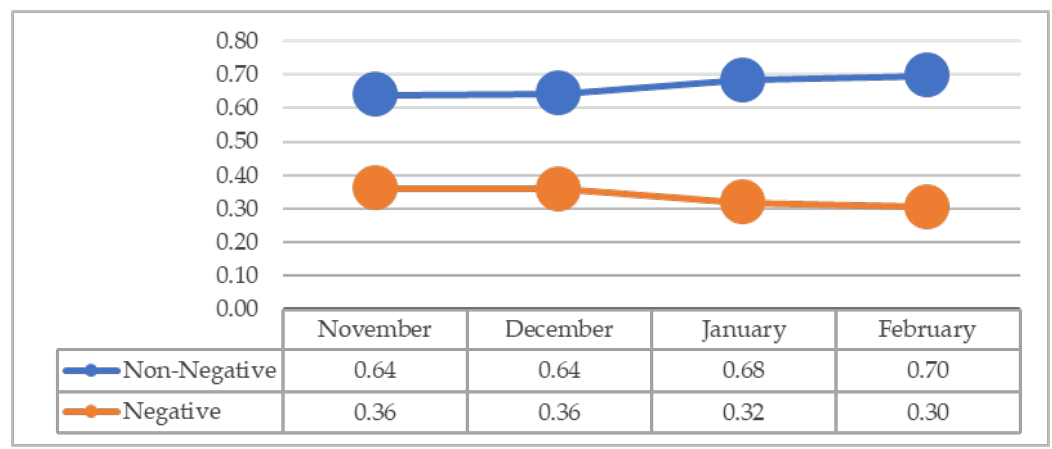

**Figure 2.** The rate of negative and non-negative tweets from November 2020 to February 2021.

The second part represents the topics discussed in the collected tweets. Table 1 shows the 26 topics in which 25 topics contained the words related to the COVID-19 vaccine. However, T24 did not show a theme related to the vaccine, indicating that some users used COVID-19 vaccine-related hashtags or keywords to amplify their other (here, sport) interests. We removed T24 for the rest of the analysis. Table 1 also shows a label and a short description for each topic, which was suggested by the human coders. The table shows a wide range of topics, such as politician hoax and vaccination management.

Figure 3 shows the measure of the average weight of each topic per tweet. This figure shows that the vaccine sites and the vaccination and election topics were the most and least popular topics. The figure also indicates that there was high discussion on topics related to vaccination hesitancy and immunity, indicating the priority of topics for Twitter users.

**Table 1.** Topics of COVID-19 related tweets and the label and short description of topics. The order of topics is based on the out of LDA.

| ID | Label | Short Description | Topic |
|---|---|---|---|
| T1 | Vaccine Exemption Bill | COVID-19 Vaccine Exemption Bill | vaccine gates people bill COVID government trust control chip forced |
| T2 | Vaccine Distribution | COVID-19 Vaccine Distribution | distribution vaccine rollout state vaccination plan COVID governor gov federal |
| T3 | Death and Vaccine | COVID-19 Death and Vaccine | COVID vaccine coronavirus news health deaths doctor cases reaction |
| T4 | Vaccine Information Sharing | COVID-19 Vaccine Information Sharing | vaccine COVID black questions information women community read vaccination vaccines public |
| T5 | Politician Hoax | Politician Hoax on COVID-19 Vaccine | vaccine line people COVID white house wait front ill hoax |
| T6 | Vaccination Sites | COVID-19 Vaccination Sites | vaccination county COVID appointments health sites mass state week clinic |
| T7 | Vaccination Hesitancy | COVID-19 Vaccination Hesitancy | vaccine people make good sense thing point understand bad science |
| T8 | Emergency Approval of Vaccines | Emergency Approval of COVID-19 Vaccines | vaccine COVID johnson fda emergency pfizer distribution approval panel moderna |
| T9 | Vaccines' Mechanism | COVID-19 Vaccines' Mechanism | vaccine flu virus immune COVID system mrna body response antibodies |
| T10 | Vaccine for Teachers | COVID-19 Vaccine for Teachers | vaccine teachers people school risk vaccination high COVID priority group |
| T11 | Vaccination, Mask, and Social Distancing | COVID-19 Vaccine, Mask, and Social Distancing | vaccine mask wear people stay social safe work virus distancing |
| T12 | Vaccine Immunity | COVID-19 Vaccine Immunity | vaccine virus immunity people rate herd COVID spread death effective |
| T13 | Vaccine Effectiveness | COVID-19 Vaccine Effectiveness | vaccine COVID pfizer effective trial data moderna test results study |
| T14 | Friends and Family Vaccination | COVID-19 Vaccination Stories of Friends and Family | vaccine COVID year family told friends mom expect wait friend |
| T15 | Vaccination and Election | COVID-19 Vaccination and Election | vaccine trump speed election plan COVID virus promised biden healthcare |
| T16 | Trump Administration Performance | Performance of Trump Administration and COVID-19 Vaccine | vaccine people lives trump americans pandemic country virus thousands dead |
| T17 | Vaccine Development Timeline | COVID-19 Vaccine Development Timeline | vaccine time months years weeks long takes work make fast |
| T18 | Vaccination for Health Workers and Nursing Home Residents | COVID-19 Vaccine for Health Workers and Nursing Home Residents | vaccine workers COVID healthcare essential nursing home residents staff hospital |
| T19 | Vaccine Management | Management of COVID-19 Vaccines | vaccine doses million COVID people week states supply shots administration |
| T20 | Travel Mandatory Testing and Vaccine | Mandatory COVID-19 Testing and Vaccine for Travel | vaccine COVID mandatory travel make proof order testing require law |
| T21 | Getting Vaccines Stories | Stories of Getting COVID-19 Vaccines | vaccine COVID today dose good shot received day tomorrow week |
| T22 | Vaccination Impact on Market | Impact of COVID-19 Vaccination on Market | vaccine end back COVID news year normal pandemic bring market |

| T23 | Biden vs. Trump on Vaccine | Biden vs. Trump on COVID-19 Vaccine | trump vaccine biden president credit administration distribution plan office team |
| --- | --- | --- | --- |
| T24 | Unrelated (Sport) | Unrelated (Sport) | vaccine game time watch back show play live dolly COVID |
| T25 | Supporting Pharmaceutical Companies | Supporting Pharmaceutical Companies for COVID-19 Vaccine Production | vaccine government COVID companies big pay make pharma funding research |
| T26 | Vaccine Side Effects | COVID-19 Vaccine Side Effects | vaccine side COVID effects shot long arm feel dose days |

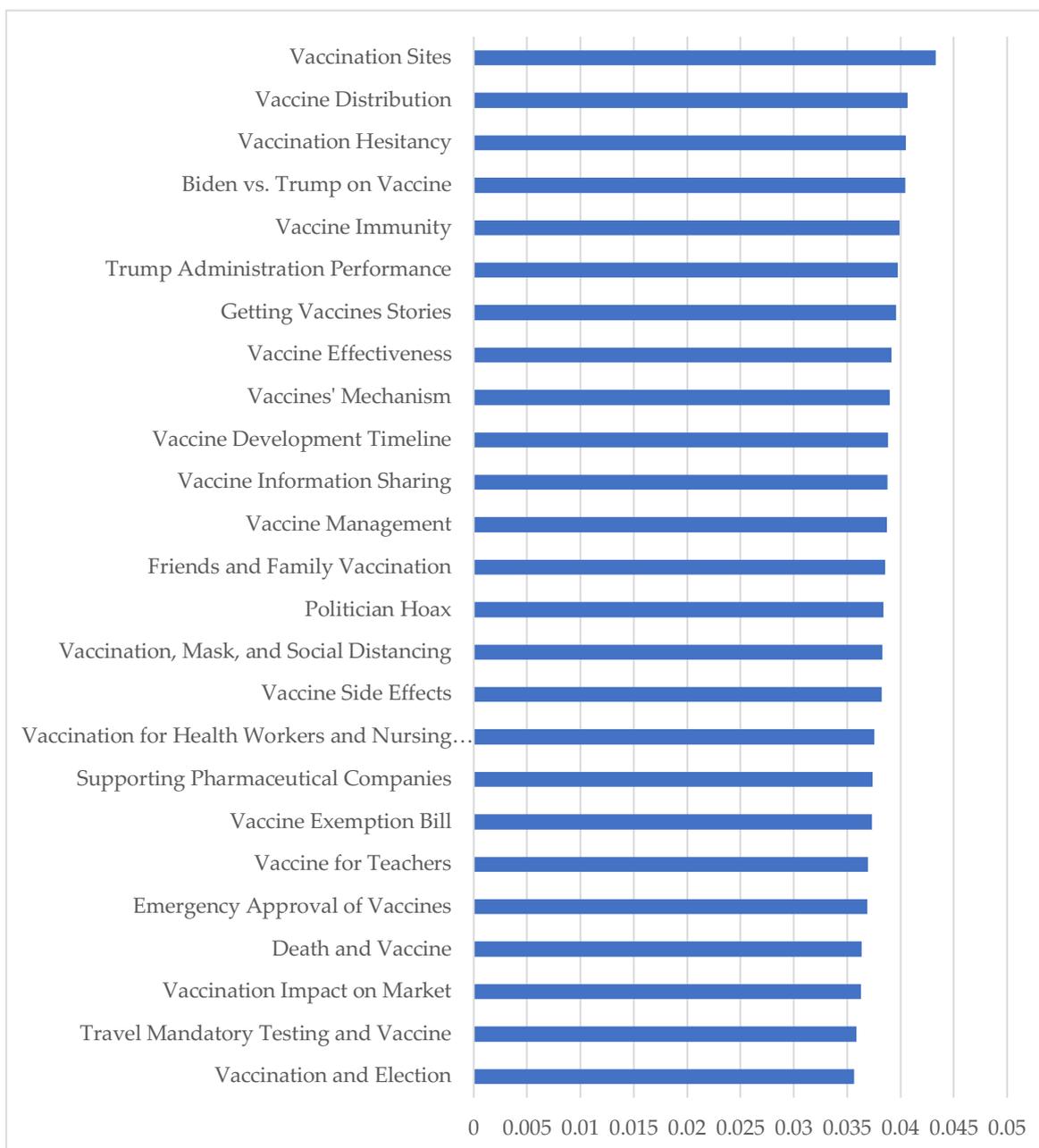

**Figure 3.** The average weight of each topic per tweet.

The third part illustrates the comparison of negative and non-negative tweets regarding the average weight of the 25 topics. Table 2 indicates that there was a significant difference (adjusted *p*-value ≤ 0.001) between negative and non-negative tweets regarding 23 topics (92%). The average weight of 12 topics (48%) was higher for the negative tweets than the non-negative ones. The rest (44%) of the 23 topics were discussed in the non-negative tweets more than the negative tweets. There was not a significant difference between negative and non-negative tweets regarding two topics: vaccine distribution and supporting pharmaceutical companies for vaccine production. Out of the significant comparisons, the effect size experiment in Table 2 indicates that the difference between the negative and non-negative tweets regarding the average weight of topics is not trivial, including one large, four medium, 13 small, and five very small effect sizes.

**Table 2.** Statistical comparison of negative and non-negative tweets regarding each topic (Neg: negative tweets; NonNeg: non-negative tweets; NS: Not Significant; *: adjusted *p*-value ≤ 0.001). The updated effect size column is based on the thresholds in Table 1.

| ID | Topic | *t*-Test Result | Cohen's d Mean | Effect Size |
|---|---|---|---|---|
| T1 | Vaccine Exemption Bill | * Neg > NonNeg | 0.2 | Small |
| T2 | Vaccine Distribution | NS | NS | NS |
| T3 | Death and Vaccine | * Neg < NonNeg | 0.2 | Small |
| T4 | Vaccine Information Sharing | * Neg < NonNeg | 0.6 | Large |
| T5 | Politician Hoax | * Neg > NonNeg | 0.3 | Small |
| T6 | Vaccination Sites | * Neg < NonNeg | 0.3 | Medium |
| T7 | Vaccination Hesitancy | * Neg > NonNeg | 0.2 | Small |
| T8 | Emergency Approval of Vaccines | * Neg < NonNeg | 0.2 | Small |
| T9 | Vaccines' Mechanism | * Neg > NonNeg | 0.2 | Small |
| T10 | Vaccine for Teachers | * Neg > NonNeg | 0.1 | Very Small |
| T11 | Vaccination, Mask, and Social Distancing | * Neg > NonNeg | 0.3 | Medium |
| T12 | Vaccine Immunity | * Neg > NonNeg | 0.2 | Small |
| T13 | Vaccine Effectiveness | * Neg < NonNeg | 0.3 | Medium |
| T14 | Friends and Family Vaccination | * Neg < NonNeg | 0.2 | Small |
| T15 | Vaccination and Election | * Neg > NonNeg | 0.2 | Small |
| T16 | Trump Administration Performance | * Neg > NonNeg | 0.4 | Medium |
| T17 | Vaccine Development Timeline | * Neg > NonNeg | 0.2 | Small |
| T18 | Vaccination for Health Workers and Nursing Home Residents | * Neg < NonNeg | 0.1 | Very Small |
| T19 | Vaccine Management | * Neg < NonNeg | 0.2 | Small |
| T20 | Travel Mandatory Testing and Vaccine | * Neg < NonNeg | 0.2 | Small |
| T21 | Getting Vaccines Stories | * Neg < NonNeg | 0.3 | Small |
| T22 | Vaccination Impact on Market | * Neg < NonNeg | 0.1 | Very Small |
| T23 | Biden vs. Trump on Vaccine | * Neg > NonNeg | 0.1 | Very Small |
| T25 | Supporting Pharmaceutical Companies | NS | NS | NS |
| T26 | Vaccine Side Effects | * Neg > NonNeg | 0.1 | Very Small |

The first topic represents discussions around legislations related to COVID-19. This topic had higher weight in negative tweets than the non-negative ones. The second topic is about the distribution of vaccines in U.S. states. We have provided examples of tweets for topics in the Supplementary file Table S1. For instance, a tweet related to T2 said, *"We can't deliver administer vaccines efficiently effectively nationwide either We never learn what to do as a fed action decentralized or privatized Pandemic was a national emergency needed a national vaccine rollout like a cat 5 hurricane hitting everywhere"* (Supplementary file Table S1). There was no significant difference between the negative and non-negative tweets regarding T2.

T3 shows tweets containing content related to the COVID-19 vaccine or death. The next topic represents tweets related to information sharing, such as, *"Need fair balance information on safety C19 vax safety Doctors should weigh decision with patients on risks and benefits C19 recovered children pregnant women lowrisk should defer"* (Supplementary file Table S1). The third and fourth topics were discussed more in non-negative tweets than the negative ones.

The fifth topic shows discussions around politicians who underestimated the COVID-19 virus but rushed to get the COVID-19 vaccine. The next topic represents tweets related to vaccine sits (e.g., working hours), like, *"Monday s vaccination appointments at Alamodome WellMed clinics rescheduled due to expected wintry weather in San Antonio KSAT San Antonio"* (Supplementary file Table S1). While the weight of T5 was higher among negative tweets than non-negative ones, the weight of T6 was higher among non-negative tweets than negative ones.

The seventh topic illustrates tweets related to vaccine hesitancy, such as *"The problem with saying this is it makes people feel the vaccine doesn't matter If you want people to get and trust the vaccine saying nothing will change if you get it or not isn't going to help in that cause"* (Supplementary file Table S1). The eighth topic shows comments discussing the approval process of COVID-19 vaccines. The negative tweets discussed the vaccine hesitancy topic more than the rest of the tweets. However, the weight of T7 was higher among non-negative tweets than negative ones.

The ninth topic represents tweets around the mechanism of COVID-19 vaccines. The tenth topic shows discussions about the vaccine priority for teachers. For example, one user posted, *"Teachers should be in a priority group for the vaccine. Virtual learning is not working for the students. I work in college admissions. These kids are not performing academically. Teachers tell me this"* (Supplementary file Table S1). Both T9 and T10 were discussed more in negative tweets than the rest of the tweets.

The next topic illustrates tweets encouraging people to wear a mask, keep social distancing, and get the vaccine, such as *"My masking does not protect me it protects you from me and your masking protects me from you. Please wash your hands wear a mask stay home if sick and get the vaccine if you can Thank you."* The twelfth topic shows posts around the immunity of COVID-19 vaccines. For instance, one user said, *"Some early evidence suggests natural immunity may not last very long. We wont know how long immunity produced by vaccination lasts until we have more data on how well the vaccines work"* (Supplementary file Table S1). Both T11 and T12 were discussed more in negative tweets than the rest of the tweets.

The thirteenth topic illustrates tweets around the efficiency of COVID-19 vaccines. The next topics shows stories of getting vaccines by family and friends, such as, *"The nursing home called me today to ask if I wanted my mother to get the COVID vaccine. Oh please please please give it to her as soon as you can. I told them I haven't hugged my mother since March. She's 96 years old and I don t know how much time I have left with her"* (Supplementary file Table S1). Both T13 and T14 were discussed more in non-negative tweets than negative tweets.

The fifteenth theme shows tweets at the intersection of the 2020 U.S. election and the COVID-19 vaccine. T16 illustrates discussions about the performance of Trump administration regarding the COVID-19 pandemic and vaccines. T15 and T16 were more popular among the negative tweets than non-negative ones.

T17 represents tweets related to the timeframe of developing the COVID-19 vaccine, such as this tweet, *"We knew it was coming but we didn't know it was going to be this year. We all though maybe next year or years from now as it usually takes 45 years to create a vaccine and then get it approved. At the start of the pandemic we never thought to see a vaccine this soon"* (Supplementary file Table S1). T18 is about the vaccine priority for health workers and nursing home residents. For example, one tweet said, *"Health care workers and nursing home residents should be at the front of the line when the first coronavirus vaccine shots become available an influential government advisory panel said Tuesday"* (Supplementary file Table S1). Both T17 and T18 are more popular among non-negative tweets than negative tweets.

T19 shows tweets regarding vaccine management. T20 illustrates discussions around mandatory test and vaccine for flying, such as *"Beginning Nov 24 anyone flying to Hawaii will have to show a negative COVID19 test before their departure no matter the airline they fly in onwhile it will require proof of vaccination to fly"* (Supplementary file Table S1). Non-negative tweets discussed T19 and T20 more than the rest of the tweets.

While T14 mostly focused on COVID-19 vaccine experiences of family and friends, T21 was more about personal experiences. For example, one user said, *"I got my appoint* *today for 2nd dose thru email from HHD COVID19 Vaccine 2nd Dose Confirmation Good news for all those waiting on appointment for 2nd dose."* The next topic shows tweets discussing the impact of the COVID-19 vaccine news on markets, such as this tweet *"Markets have spent November celebrating upbeat vaccine news and closure on US election uncertainty After a strong month are equities headed for another reset"* (Supplementary file Table S1). Both T14 and T15 were discussed more in non-negative tweets than non-negative ones.

T23 is about vaccine management by Trump vs. Biden. T25 shows the opinion of users on supporting pharmaceutical companies. For example, one user posted the following tweet, *"He signed the national vaccine injury compensation bill that exempts vaccine makers from liability for childhood vaccinations No surprise there"* (Supplementary file Table S1). The final topic represents tweets sharing the side effect information of COVID-19 vaccines. While there was no significant difference between negative and non-negative tweets regarding the weight of T25, T23 and T26 were discussed more in negative tweets than the rest of the tweets.

Table 3 shows the top-5 topics of negative and non-negative tweets based on the weight of topics. This table shows that there are mostly negative topics regarding the Trump administration, vaccination hesitancy, vaccine immunity, vaccine, mask, and social distancing, and comparing vaccination strategies of presidents Biden and Trump. On the other side, Twitter users showed the most non-negative attitude toward vaccination sites, stories of getting vaccines, and vaccine effectiveness, information, and management. There is no common topic between the top-5 topics of negative and non-negative tweets.

**Table 3.** Top-5 Topics of Negative and Non-Negative Tweets.

| Negative Tweets | Non-Negative Tweets |
| --- | --- |
| Trump Administration Performance | Vaccination Sites |
| Vaccination Hesitancy | Stories of Getting Vaccines |
| Vaccine Immunity | Vaccine Effectiveness |
| Vaccination, Mask, and Social Distancing | Vaccine Information |
| Biden vs. Trump on Vaccine | Vaccine Management |

## 4. Discussion

Understanding public opinion is an important process in public health, which is a labor-intensive and time-consuming process. It is necessary to develop an efficient framework to explore public opinion. This research attempted to develop a systematic and objective research framework as much as possible, which lead to interesting results. This study addressed the following four research questions: (1) How did the sentiment of tweets related to the COVID-19 vaccine change between November 2020 and February 2021? (2) What are the main topics in tweets related to the COVID-19 vaccine?, (3) Is there a significant difference between topics in negative and non-negative tweets?, and (4) What are the top topics in negative and non-negative tweets? We found that the negative and non-negative sentiment of tweets containing terms related to the COVID-19 vaccine had decreasing and increasing trends, respectively. We identified 25 topics in our corpus and found that there was a significant difference between negative and non-negative tweets regarding the weight of most topics. Finally, our final analysis shows that vaccination sites, stories of getting vaccines, and vaccine effectiveness, information, and management were top-5 topics of negative tweets and Trump administration, vaccination hesitancy, vaccine immunity, vaccine, mask, and social distancing, and comparing vaccination strategies of presidents Biden and Trump were top-5 topics of non-negative tweets.

This paper investigated how public sentiment changed on Twitter during the last two months of 2020 and the first two months of 2021 when people discussed vaccine trials, approval, and distribution. Our findings show that the negative sentiment regarding the COVID-19 vaccine had a decreasing trend between November 2020 and February 2021. The vaccination in the U.S. was started on 14 December 2020. This result indicates that U.S. public sentiment has become less negative during the two months after starting the vaccination. This result is in line with a survey indicating that more people are more willing to get the vaccine from November 2020 to February 2021 [64] and a research study showing mostly positive emotions in tweets posted between January and October 2020 [35].

We found that Twitter users have discussed a wide range of topics from vaccination sites to vaccination and election between November 2020 and February 2021. Topic modeling showed that the first two topics were about vaccination sites and distributions. This illustrates that the public would like to know how to get the vaccine. In January and February 2021, more vaccines were available, and more people got the vaccine. Some topics show the experiences and stories of people and their friends and family who got the COVID-19 vaccine. Twitter users talked about political issues such as comparing Trump and Biden. Some topics are issues related to vaccines, their performance, immunity, mechanism, development timeline, side effects, and approval process. While relevant studies focused on few general topics including attitudes [35], vaccine development [35], complaints in Australia [35], and vaccine hesitancy [65] and were developed mostly on tweets posted in 2020 before starting the vaccination in the U.S., this paper identified a wide range of topics posted in four months in 2020 and 2021 and provided more details on issues discussed on Twitter. According to a survey [64], side effects, emergency approval, and effectiveness are the main concerns of people who are reluctant to get the COVID-19 vaccine. This study identified not only these concerns but also other topics.

To the best of our knowledge, this is the first study compared negative and non-negative tweets related to the COVID-19 vaccine and identified top topics of those tweets posted in U.S. Our results illustrate that the vaccine discussions on Twitter are evolving, with negative and non-negative attitudes on different issues. This result indicates that the negative and non-negative tweets had different priorities and focuses.

*4.1. Strengths and Implications*

While traditional health surveys are limited, this study suggests that using social media is efficient and can complement the surveys. Social media may provide a more insightful approach to health management. This research can be utilized for not only COVID-19 vaccine but also other health issues. This research is beneficial to the science of engagement to monitor public opinion regarding health issues and offers a vision for developing interventions. This study can also help health organizations and practitioners to effectively and efficiently use social media for population monitoring, strengthen communication, and empowering policymaking.

Social media can be utilized to facilitate health information sharing and to inform health agencies on emerging population-level trends. Our findings have implications for health communication through health campaigns and the news media. According to a survey, more than 1 billion worldwide would refuse a free vaccine, ranging from 4% in Myanmar to 75% in Kazakhstan [66]. One practical example is that, because vaccine immunity concern is among the top-5 topics of negative tweets, it is reasonable for news media to provide more information and health campaigns to develop interventions targeting that issue.

The topics that have a higher weight among negative tweets could be considered more than other topics in developing health monitoring and communication strategies. Public health campaigns can include more information or develop new strategies about the side effects of COVID-19 vaccines. Like the work accomplished by Moehring et al. [26], we also suggest using factual normative messages and different interventions to reduce perceived barriers.

*4.2. Limitations and Future Work*

Considering the limitations of this study can provide new opportunities for future work. First, Twitter data constitutes biases. The first one is that Twitter is not a representative sample. Twitter is used by less than 25% of U.S. adults, mostly democrats aged between 18 and 49 years old, who have higher incomes, and are more educated than the average person in the general public [67]. The next bias is that non all Twitter users post comments about COVID-19 vaccines. Our study is limited to the users who share opinion regarding COVID-19 vaccines. The third bias is that Twitter users share content that they feel comfortable, which might not represent their true opinion and feeling. The fourth one is that there might be factors that can affect the results but were not considered in this study.

Second, there is a lack of demographic information (e.g., race) regarding each user. Third, our data collection was limited to a few queries, indicating that we might have missed many other relevant tweets. Thus, our findings should be interpreted as suggestive of sentiment and semantic trends rather than strong evidence of them.

Another limitation of social media analysis is that the trends are usually very sensitive to specific non-systematic events and news. For example, when a public figure such as the president speaks about the pros and cons of the vaccine, there might be a surge in the number of tweets that discuss vaccination. Although taking this into account can lead to more insights into public opinion, the way, in which we have analyzed the Twitter data over the whole period of this study, does not suffer much from sudden ups and downs in the number of tweets corresponding to such public events.

Despite these limitations, this study can serve as a baseline approach for developing and enhancing similar research. In addition, our results are obtained from a new, large, and non-traditional dataset and can provide new insights into discussions related to the COVID-19 vaccine. Future work could address the limitations of this research by using inferring methods to predict demographic information of users and incorporating other social media platforms (e.g., Reddit).

There are several paths through which this research can become more proliferated. For example, previous research has shown that demographic characteristics, political affiliation, age, income level, and college education are good predictors of vaccine hesitancy [27,68]. Where possible, the trends over time data for topics can be stratified to account for different demographic characteristics, to see whether if a meaningful difference in the trends would also exist in the Twitter data, that corresponds with user demographic characteristics. In addition, the number of tweets per time unit within the total period of this study can also be indicative of a few things, such as the perceived importance of the vaccination for Twitter users.

## 5. Conclusions

From a methodological perspective, the results show that Twitter can give suggestions about sentiment and semantic trends to the COVID-19 vaccine, although the limitations mean that they must be interpreted cautiously. Twitter is the only practical source of large-scale content from the public to develop this type of study. In theory, it would be possible to apply similar larger-scale analyses on data from social media platforms sources, such as Facebook.

**Supplementary Materials:** The following are available online at www.mdpi.com/xxx/s1. Supplementary Table S1. Topics and Examples of Tweets.

**Author Contributions:** Conceptualization, A.K.; methodology, A.K.; software, A.K.; validation, A.K., M.Z., B.G., and H.R.B.; formal analysis, A.K.; investigation, A.K., M.Z., B.G., H.R.B., and M.M.N.; resources, A.K.; data curation, A.K.; writing—original draft preparation, A.K., M.Z., B.G., H.R.B., and M.M.N.; writing—review and editing, A.K., M.Z., B.G., H.R.B., and M.M.N.; visualization, A.K.; supervision, A.K.; project administration, A.K.; funding acquisition, A.K. All authors have read and agreed to the published version of the manuscript.

**Funding:** This research was partially supported by the Big Data Health Science Center (BDHSC) at the University of South Carolina. All opinions, findings, conclusions, and recommendations in this paper are those of the authors and do not necessarily reflect the views of the funding agency.

**Institutional Review Board Statement:** Not applicable.

**Informed Consent Statement:** Not applicable.

**Conflicts of Interest:** The authors declare no conflict of interest.